\documentclass[11pt]{article}

\usepackage{amsfonts,amssymb,amsthm,amsmath}
\usepackage{hyperref}
\usepackage{mathtools}
\usepackage{comment}
\usepackage[margin=1in]{geometry}

\usepackage{algorithm}
\usepackage{algpseudocode}
\usepackage{accents}

\usepackage{bm}

\usepackage{fancybox}
\usepackage{color}
\usepackage{latexsym}

\usepackage{eepic}
\usepackage{epic}
\usepackage{epsf}
\usepackage{graphics}

\newtheorem{theorem}{Theorem}
\newtheorem*{theorem*}{Theorem}

\newtheorem{lemma}[theorem]{Lemma}

\newtheorem{fact}[theorem]{Fact}

\newtheorem{claim}[theorem]{Claim}

\newtheorem{proposition}[theorem]{Proposition}
\theoremstyle{definition}

\newtheorem{definition}[theorem]{Definition}
\newtheorem{remark}[theorem]{Remark}
\newcommand{\B}{\mathfrak{B}}
\newcommand{\A}{\mathcal{A}}
\newcommand{\M}{\mathbb{M}}
\renewcommand{\r}{\mathfrak{r}}

\renewcommand{\H}{\mathcal{H}}

\newcommand{\F}{\mathbb{F}}
\newcommand{\G}{\mathcal{G}}

\newcommand{\Q}{\mathbb{Q}}

\DeclareMathOperator{\poly}{poly}

\DeclareMathOperator{\Gal}{Gal}

\DeclareMathOperator{\diag}{diag}
\DeclareMathOperator{\dom}{dom}

\DeclareMathOperator{\cir}{circ}

\renewcommand{\ge}{\geqslant}
\renewcommand{\geq}{\geqslant}
\renewcommand{\le}{\leqslant}
\renewcommand{\leq}{\leqslant}

\DeclareFontFamily{OMX}{MnSymbolE}{}
\DeclareFontShape{OMX}{MnSymbolE}{m}{n}{
   <-6>  MnSymbolE5
   <6-7>  MnSymbolE6
   <7-8>  MnSymbolE7
   <8-9>  MnSymbolE8
   <9-10> MnSymbolE9
   <10-12> MnSymbolE10
   <12->   MnSymbolE12}{}
\DeclareSymbolFont{mnlargesymbols}{OMX}{MnSymbolE}{m}{n}
\SetSymbolFont{mnlargesymbols}{bold}{OMX}{MnSymbolE}{b}{n}
\DeclareMathDelimiter{\llangle}{\mathopen}{mnlargesymbols}{'164}{mnlargesymbols}{'164}
\DeclareMathDelimiter{\rrangle}{\mathclose}{mnlargesymbols}{'171}{mnlargesymbols}{'171}

\renewcommand{\angle}[1]{\langle #1 \rangle}
\newcommand{\ubar}[1]{\underaccent{\bar} #1}
\newcommand{\dangle}[1]{{\llangle} #1 {\rrangle}}

\DeclareFontEncoding{LS1}{}{}
\DeclareFontSubstitution{LS1}{stix}{m}{n}
\DeclareSymbolFont{symbols2stix}{LS1}{stixfrak} {m} {n}
\DeclareMathSymbol{\lparenless}{\mathopen} {symbols2stix}{"32}
\DeclareMathSymbol{\rparengtr}{\mathclose}{symbols2stix}{"33}

\newcommand{\newbrak}[1]{{\lparenless} #1 {\rparengtr}}
\newcommand{\sm}{\mbox{\small\rm SM}}

\newcommand{\maintheorem}{\noindent{\it Proof of Theorem~\ref{main-theorem}.}}

\title{}
\date{}

\title{Black-box Identity Testing of Noncommutative Rational Formulas of Inversion Height Two in Deterministic Quasipolynomial-time}
\author{
V. Arvind\thanks{Institute of Mathematical Sciences (HBNI), Chennai, India, \texttt{email: arvind@imsc.res.in}}  
\and Abhranil Chatterjee\thanks{Indian Institute of Technology Bombay, India, \texttt{email: abhneil@gmail.com}} 
\and Partha Mukhopadhyay\thanks{Chennai Mathematical Institute, Chennai, India,\texttt{email: partham@cmi.ac.in}}
}
\begin{document}

\maketitle
\begin{abstract}
Hrube\v{s} and Wigderson \cite{HW15} initiated the complexity-theoretic study of noncommutative formulas with inverse gates. They introduced the Rational Identity Testing (RIT) problem which is to decide whether a noncommutative rational formula computes zero in the free skew field. In the white-box setting, deterministic polynomial-time algorithms are known for this problem following the works of  Garg, Gurvits, Oliveira, and Wigderson~\cite{GGOW16} and Ivanyos, Qiao, and Subrahmanyam~\cite{IQS18}.

A central open problem in this area is to design efficient deterministic \emph{black-box} identity testing 
algorithm for rational formulas. In this paper, we solve this for the first nested inverse case. More precisely, we obtain a deterministic quasipolynomial-time black-box RIT algorithm for noncommutative rational formulas of inversion height two via a hitting set construction. 
Several new technical ideas are involved in the hitting set construction, including key concepts from matrix coefficient realization theory~\cite{vol18} and properties of cyclic division algebra~\cite{Lam01}. En route to the proof, an important step is to embed the hitting set of Forbes and Shpilka for noncommutative formulas~\cite{FS13} inside a cyclic division algebra of small index.
\end{abstract}

\thispagestyle{empty}


\newpage
\section{Introduction}\label{sec:intro}
The broad goal of \emph{algebraic complexity} is to study the complexity of computing polynomials and rational functions using basic arithmetic operations:  additions, multiplications, and inverses. \emph{Arithmetic circuits} and \emph{arithmetic formulas} are two extensively studied models of computation. An important sub-area of algebraic complexity is the \emph{noncommutative computation} where the variables do not commute (i.e. $xy\neq yx$). If we allow only the addition and multiplication gates in the noncommutative formulas/circuits, they compute noncommutative polynomials (similar to the commutative case) in the free algebra.

In the commutative case, the role of inverses is well understood, but in the noncommutative world it is quite subtle. To elaborate, it is known that \emph{any} commutative rational expression can be expressed as $fg^{-1}$ where $f$ and $g$ are two commutative polynomials~\cite{Str73}. However, noncommutative rational expressions (formulas with inverses) such as $x^{-1} + y^{-1}$ or $xy^{-1}x$ cannot be represented as $f g^{-1}$ or $f^{-1} g$. If we have \emph{nested inverses}, it makes the rational expression more complicated, for example ${(z + xy^{-1}x)}^{-1}- z^{-1}$. 
Moreover, a noncommutative rational expression is not always defined on a matrix substitution. For a noncommutative rational expression $\r$, its \emph{domain of definition} is the set of matrix tuples (of any dimension) where $\r$ is defined. We denote it by $\dom(\r)$. Two rational expressions $\r_1$ and $\r_2$ are \emph{equivalent} if they agree on $\dom(\r_1)\cap \dom(\r_2)$. This induces an equivalence relation on the set of all noncommutative rational expressions (with nonempty domain of definition). It was used by Amitsur in his characterization of the \emph{universal} free skew field \cite{ami66} and the equivalence classes are called \emph{noncommutative rational functions}. 

The \emph{inversion height} of a rational formula is the maximum number of inverse gates in a path from an input gate to the output gate. It is known \cite{HW15} that the inversion height of a rational formula of size $s$ is bounded by $O(\log s)$. 
Hrube\v{s} and Wigderson \cite{HW15} consider the \emph{rational identity
testing} problem (RIT) of testing the equivalence of two rational formulas. It is the same as testing whether a rational formula is computing the zero function in the free skew field. In other words, decide whether there exists a matrix tuple (of any dimension) such that the rational formula evaluates to nonzero on that substitution. Rational expressions exhibit peculiar properties which seems to make the RIT problem quite different from polynomial identity testing. For example, Bergman has constructed an explicit rational expression, of inversion height two, which is an identity for $3\times 3$ matrices but not an identity for $2\times 2$ matrices \cite{Breg76}.  Also, the apparent lack of \emph{canonical representations}, like sum of monomials representation for polynomials, and the use of nested inverses in noncommutative rational expressions complicate the problem. For example, the rational expression $(x+xy^{-1}x)^{-1} + (x+y)^{-1} -x^{-1}$ of inversion height two is a rational identity, known as Hua's identity~\cite{Hua49}.

However, Hrube\v{s} and Wigderson give an efficient reduction from the RIT problem to the singularity testing problem of linear pencils.  
A \emph{linear pencil} $L$ of size $s$ over noncommuting variables $\ubar{x}=\{x_1, \ldots, x_n\}$ is a $s\times s$ matrix whose entries are linear forms in $\ubar{x}$ variables, i.e.\ $L = A_0 + \sum_{i=1}^n A_i x_i$,  where each $A_i$ is an $s\times s$ matrix over the field $\F$. A rational function $\r$ in $\F\newbrak{\ubar{x}}$ has a \emph{linear pencil representation} $L$ of size $s$, if for some $i,j \in [s]$, $\r =  (L^{-1})_{i,j}$. In particular, if $\r$ is  a rational formula of size $s$, Hrube\v{s} and Wigderson have shown that $\r$ has a linear pencil representation $L$ of size at most $2s$ such that $r$ is defined on a matrix tuple if and only if $L$ is invertible on that tuple~\cite{HW15}. 
Using this connection, they reduce the RIT problem to the problem of testing whether a given linear pencil is invertible over the free skew field in deterministic polynomial time. 
The latter one is the noncommutative SINGULAR problem, whose commutative analogue is the symbolic determinant identity testing problem. The deterministic complexity of symbolic determinant identity testing is completely open \cite{KI04} in the commutative setting.  In contrast, the SINGULAR problem in noncommutative setting has deterministic polynomial-time algorithms in the white-box model due to \cite{GGOW16,IQS18}. The algorithm in \cite{GGOW16} is based on operator scaling and the algorithm in \cite{IQS18} is based on the second Wong sequence and a constructive version of \emph{regularity lemma}. As a consequence, a deterministic polynomial-time white-box RIT algorithm follows. 

A central open problem is to design an efficient \emph{deterministic} RIT algorithm in the black-box case \cite{GGOW16}. There is a randomized polynomial-time black-box algorithm for the problem \cite{DM17}. Can we derandomize this result even in some restricted setting, for example when the inversion height of the input rational formula is small?  
Notice that inversion height zero rational formulas are just noncommutative formulas, and a result of Forbes and Shpilka have shown a deterministic quasipolynomial-time identity testing for those (more generally, for noncommutative ABPs) via a hitting set construction \cite{FS13}. Whether their approach can be extended to the RIT problem for rational formulas is a natural direction and we prove the following theorem which is our main result. 



\begin{theorem}\label{main-theorem}
For the class of rational formulas in $\Q\newbrak{x_1, \ldots, x_n}$ of inversion height two and size at most $s$, we can construct a hitting set ${\mathcal{H}}\subseteq \M^n_{d}(\Q)$ of size $(ns)^{O(\log ns)}$ in deterministic $(ns)^{O(\log ns)}$-time. The parameter $d$ is $\poly(s,n)$ bounded.  
\end{theorem}



Prior to this work, no such hitting set construction was known that could handle nested inverses.
As we discuss in the next section, even to derandomize RIT for the special case of inversion height two, we need to accumulate several ideas involving cyclic division algebras \cite{Lam01} and matrix coefficient realization theory \cite{vol18} combined with the hitting set construction in \cite{FS13}. 


\subsection*{Proof Idea} 
Consider the following noncommutative rational formula, $\r= [x,y]^{-1}=(x y - y x)^{-1}$. Clearly there is no point in $\dom(\r)$ from the ground field, and the natural idea is to expand the series around a matrix point. Let $(p_1, p_2)$ be a matrix pair such that $[p_1, p_2]$ is invertible and let $\r(p_1, p_2)=[p_1, p_2]^{-1} = q$. Then, 
\[
\r(x + p_1, y+ p_2 ) = \left( [p_1, p_2] - [p_2, x] - [y, p_1] - [y,x]\right)^{-1}. 
\] 
Simplifying this we can write $\r(x+p_1, y+p_2) = (I -g(x,y))^{-1} q$ where $g(x,y)=q ([p_2, x] 
+ [y, p_1] + [y,x])$. Now expanding this using 
$(I - g(x,y))^{-1}=\sum_{i\geq 0} (g(x,y))^{i}$, we can see that every term in the expansion looks like $a_0 z_1 a_1 z_2 \ldots a_{d-1} z_d a_{d}$ where 
each $a_j$ is a matrix and $z_j\in\{x,y\}$. In the language of matrix coefficient realization theory \cite{vol18}, such terms (resp. series) are called generalized words or monomials (resp. generalized  series). 
In fact if a rational formula $\r$ of size $s$ has a defined point $\ubar{u}$ in some  dimension $l$ (in other words $\ubar{u} \in \dom(\r)$, and we use it interchangeably), Vol\v{c}i\v{c} shows that one can associate a special class of generalized series, a recognizable generalized series to the shifted rational formula \cite{vol18}:
\[
\r(\ubar{x} + \ubar{u}) = \boldsymbol{c} \left(I_{2ls} - \sum_{j=1}^n A^{x_j}\right)^{-1}\boldsymbol{b}.
\]
Here $\boldsymbol{c}\in(\M_{l}(\F))^{1\times 2s}$ and $\boldsymbol{b}\in(\M_{l}(\F))^{2s\times 1}$. 
The matrices $A^{x_j}, 1\leq j\leq n$ are of dimension $2s\times 2s$ as a block matrix and 
$(k_1,k_2)^{th}$ entry of $A^{x_j}$ is given by a generalized linear form $C_{k_1, k_2,j} x_j C'_{k_1,k_2,j}$ where $C_{k_1, k_2,j}, C'_{k_1,k_2,j}\in \M_{l}(\F)$. 

Focusing on our problem for rational formulas of inversion height two, the first step is to construct a quasipolynomial-size set $\H_1$ of matrix tuples of small dimension such that for every nonzero rational formula $\r$ of inversion height two, there exists a point $\ubar{u}\in\H_1$ on which $\r$ is defined. Given such a point, testing whether $\r$ is zero or not reduces to testing whether the generalized series $\r(\ubar{x} + \ubar{u})$ is zero or not. This is formally stated in Theorem \ref{rit-connection}. For a recognizable series in algebraic automata theory, a standard result by Sch\"{u}tzenberger shows that the identity testing of such infinite series is equivalent to the identity testing of 
polynomial obtained by truncation of the series up to a small degree \cite[Corollary 8.3]{Eilenberg74}. We can adapt this result in the case of generalized series too and observe that the truncated generalized polynomial (of small degree $d$) can be represented by an algebraic branching program with edge labels are linear forms over matrices. Such ABPs can be identity tested efficiently using an adaptation of the hitting set construction shown by Forbes-Shpilka \cite{FS13}. 

Although it is not clear how to carry out the truncation in the black-box setting, we are able to show that a suitable scaling of the hitting set for such generalized ABPs are good enough to hit the generalized series too. To fit the dimension correctly, throughout the computation the coefficient matrices should be embedded in the matrix algebra of dimension $dl$ using the inclusion map 
$\iota : a \rightarrow a \otimes I_d$. This is shown in Proposition \ref{prop-gen-Sch}.   
    
Clearly, $\r$ is defined at a point $\ubar{u}$ if and only if all the maximal sub-formulas of inversion height one in $\r$ evaluate to invertible matrices on $\ubar{u}$. One can consider the product of all such maximal formulas and thus our goal is now re-defined: construct 
$\H_1$ such that for every size-$s$ rational formula $\r$ of inversion height one, there is a point $\ubar{u}\in\H_1$ at which $\r(\ubar{u})$ is \emph{invertible}. We call such a hitting set a \emph{strong hitting set}. A rational formula $\r$ of inversion height one is defined at a point $\ubar{v}$ if and only if all sub-formulas which are input to inverse gates evaluate to invertible matrices on $\ubar{v}$. These sub-formulas are just noncommutative formulas. Since the Forbes-Shpilka hitting set \cite{FS13} for noncommutative formulas consists of tuples of nilpotent matrices, it is not directly applicable to our problem. 

However, it is possible to adapt their construction and get a strong hitting set, also of quasipolynomial size, such that every size-$s$ nonzero noncommutative formula evaluates to an invertible matrix on some matrix tuple in the strong hitting set \footnote{This was first explicitly constructed in \cite{ACDM20}.}. In particular, all matrices in the hitting set construction will be invertible and have the following shape: 
\[
\begin{bmatrix}
    0       &* & 0 & \cdots & 0 \\
     0       & 0 &* &\cdots  &  0 \\
      \vdots & \vdots &\ddots &\ddots  &  \vdots \\
      0       & 0 &\cdots                  & 0 &* \\
      *       & 0 & \cdots & 0 & 0
\end{bmatrix},
\]
where the dimension is determined by the depth of the noncommutative formulas.  
Expanding $\r$ around such a point would again lead to a generalized series, and (a somewhat more involved) truncation and scaling argument shows that we can get a strong hitting set for $\r$ by constructing a strong hitting set for  \emph{generalized} ABPs whose edges are labeled by linear forms over matrices. This is the essence of the second part of Proposition \ref{prop-gen-Sch}.  

At this point we face a serious obstacle. How do we find invertible matrices in the image of the generalized ABPs? In other words, how to construct a strong hitting set for generalized ABPs? The main insight is that, if the matrices present in the linear forms of the generalized ABPs are from a division algebra, then one can construct a strong hitting set from a hitting set. To implement this, we construct the hitting set for noncommutative formulas (which are of inversion height zero) over a division algebra of small index, and expand the rational formula with respect to the points in that hitting set. Why does it work? Roughly speaking, as already mentioned it is easier to find a nonzero in the image of generalized ABPs and if the computation occurs inside a division algebra then a computed nonzero element is also invertible.  

Section \ref{sec:genabp-hs} elaborates on this idea. In particular, Lemma \ref{ref-lemma-divalg-witness} provides an existential argument showing that if the linear forms of the generalized ABP are defined over a division algebra of dimension $\ell$, then there \emph{exits} a substitution to the variables from $D$ such that the generalized ABP evaluates to an invertible matrix. The proof uses two ideas. Firstly, we show that such a point exists inside the full matrix algebra of dimension $\ell$. Then we use Proposition \ref{full-space} to find such a certificate inside $D$. Once we establish the existential argument, we can use a reduction to the hitting set construction of ROABPs (in unknown order) \cite{AGKS15} to construct the hitting set in quasi-polynomial time.  
To work out the technical details we need to employ the inclusion map $\iota': a\rightarrow I_d \otimes a$ for the coefficients which are now elements of division algebra. In ring theory the maps $\iota$ and $\iota'$ are compatible: by the Skolem-Noether theorem \cite[Theorem 3.1.2]{row80} there is an invertible matrix $q_0$ such that $q_0 (I_d \otimes a) q^{-1}_0 = a\otimes I_d $ for all $a$. However, in our case we give a simple explicit construction of a permutation matrix $q_0$. 

In the remaining part of the proof sketch, we informally describe how to find a hitting set for noncommutative formulas (more generally for noncommutative ABPs) in a division algebra of small index. For simplicity, suppose the ABP degree is $2^d$. The Forbes-Shpilka hitting set \cite{FS13} has a recursive construction and it is by a reduction to the hitting set construction for ROABPs (read-once algebraic branching programs) over the commutative variables $u_1, u_2, \ldots, u_{2^d}$. The recursive step in the construction is by combining of hitting sets (via hitting set generator $\mathcal{G}_{d-1}$) for two halves of degree $2^{d-1}$ \cite{FS13} with a rank preserving step of matrix products to obtain the generator $\mathcal{G}_d$ at the $d^{th}$ step. More precisely, $\G_d$ is a map from $\F^{d+1}\rightarrow \F^{2^{d}}$ that stretches the seed $(\alpha_1, \ldots, \alpha_{d+1})$ to a $2^d$ tuple for the read-once variables.  

For our purpose, we take a classical construction of cyclic division algebras  \cite[Chapter 5]{Lam01}. The division algebra $D=(K/F,\sigma,z)$ is defined using a indeterminate $x$ as the $\ell$-dimensional vector space:
\[
D = K\oplus Kx\oplus \cdots \oplus Kx^{\ell-1},
\]
where the (noncommutative) multiplication for $D$ is defined by
$x^\ell = z$ and $xb = \sigma(b)x$ for all $b\in K$.
Here $\sigma : K\rightarrow K$ is an automorphism of the Galois group $\Gal(K/F)$. The field $F=\Q(z)$ and $K=F(\omega)$, where $z$ is an indeterminate and $\omega$ is an $\ell^{th}$ primitive root of unity. The matrix representation of a general element in $D$ is of the following form: 
$$
\begin{bmatrix}
    0       & b & 0 & \cdots & 0 \\
     0       & 0 & \sigma(b) &\cdots  &  0 \\
      \vdots & \vdots &\ddots &\ddots  &  \vdots \\
      0       & 0 &\cdots                  & 0 &\sigma^{\ell-2}(b) \\
      z\sigma^{\ell-1}(b)       & 0 & \cdots & 0 & 0
\end{bmatrix}.
$$

To embed the hitting set of \cite{FS13}, we need to choose $\ell=2^{L}$ appropriately larger than $2^d$. As it turns out the construction of the division algebra requires a tower of extension fields of $F$, with a higher order root of unity at each stage. 

Specifically, let $\omega_i =\omega^{2^{a_i}}$ for $a_1>a_2>\cdots > a_d > 0$,
where $a_i$ are positive integers suitably chosen. Let $K_i=F(\omega_i)$
be the cyclic Galois extension for $1\le i\le d$ giving a tower
of extension fields
\[
F\subset F(\omega_1) \subset F(\omega_2)\subset \cdots \subset F(\omega_d) \subset F(\omega).
\]

As we show in Section \ref{sec:embedding} that 
we require two properties of $\omega_i, 1\le i\le d$.
Firstly, for the hitting set generator $\G_i$ we will choose the root of
  unity as $\omega_i$ and the variable $\alpha_i$ will take values
  only in the set $W_i = \{\omega_i^j\mid 1\le j\le 2^{L-a_i}\}$.
We also require that the $K$-automorphism $\sigma$ has the property
  that for all $1\le i\le d$ the map $\sigma^{2^i}$ fixes
  $\omega_i$. In fact we will ensure that $\sigma^{2^i}$ has
  $F(\omega_i)$ as its fixed field.
The construction of $D$ satisfying the above properties is the main technical step in Section \ref{sec:embedding}. 

Implementing all these steps we get a quasipolynomial-size hitting set over $\Q(\omega,z)$. Then we show how to transfer the hitting set over $\Q$ itself by a relatively standard idea that treats the parameters $\omega$ and $z$ as \emph{fresh} \emph{indeterminates} $t_1,t_2$ and vary them over a suitably chosen polynomial-size set. This is sketched in Section \ref{sec:all-together}.

One may naturally wonder whether our proof technique can be extended to higher inversion heights. We include a brief discussion about this in Section \ref{sec:conclusion}.

\subsection*{Organization} 
In Section \ref{sec:prelim}, we collect some background results from algebraic complexity theory, matrix coefficient realization theory, and cyclic division algebra. Section \ref{sec:embedding} contains the proof that the Forbes-Shpilka hitting set can be embedded in a cyclic division algebra of small index. In Section \ref{sec:genabp-hs}, we construct a quasipolynomial-size strong hitting set for generalized ABPs over division algebra. Finally, in Section \ref{sec:all-together} we combine the results developed in Section \ref{sec:embedding} and Section \ref{sec:genabp-hs} to obtain our main result which gives a quasipolynomial-size hitting set for rational formulas of inversion height two. In Section \ref{sec:conclusion}, we mainly discuss the possibility of extending our method to higher inversion heights.     


\section{Background and Notation}\label{sec:prelim} 

Throughout the paper, we use $\F, F, K$ for fields. The notation $\M_m(\F)$ (respectively, $\M_m(F), \M_m(K)$) are used for $m$ dimensional matrix algebra over $\F$ (respectively over $F, K$) where $m$ is clear from the context. $D$ is used to denote cyclic division algebras. 
Let $\ubar{x}$ be the set of variables $\{x_1, \ldots, x_n\}$. Sometime we use notation like $\ubar{u}, \ubar{v}, \ubar{p}, \ubar{q}$ to denote the matrix tuples in suitable matrix algebras. The free noncommutative ring of polynomials over a field $\F$ is denoted by $\F\angle{\ubar{x}}$. The ring of \emph{formal power series} is denoted by $\F\dangle{\ubar{x}}$. 
For a series (or polynomial) $S$, the coefficient of a monomial (word) in $S$ is denoted by $[m]S$. 


\subsection{Algebraic Complexity} 


\begin{definition}[Algebraic Branching Program]\label{abpdefn}
An \emph{algebraic branching program} (ABP) is a layered directed
acyclic graph. The vertex set is partitioned into layers
$0,1,\ldots,d$, with directed edges only between adjacent layers ($i$
to $i+1$). There is a \emph{source} vertex of in-degree $0$ in layer
$0$, and one out-degree-$0$ \emph{sink} vertex in layer $d$. Each edge
is labeled by an affine $\F$-linear form. The polynomial computed by
the ABP is the sum over all source-to-sink directed paths of the
ordered product of affine forms labeling the path edges. 
\end{definition}

The \emph{size} of the ABP is defined as the total number of nodes and the \emph{width} is the maximum number of nodes in a
layer.  The ABP model is defined for computing commutative or
noncommutative polynomials. ABPs of width $r$ can also be seen as
iterated matrix multiplication $ \boldsymbol{c}\cdot M_1 M_2 \cdots
M_{\ell} \cdot\boldsymbol{b} $, where $\boldsymbol{c}, \boldsymbol{b}$ are
$1\times r$ and $r \times 1$ vectors respectively and each $M_i$ is a $r \times r$ matrix, whose
entries are affine linear forms over $\ubar{x}$.

We also consider commutative set-multilinear ABPs and read-once oblivious ABPs (ROABPs). For the set-multilinear case, the
(commutative) variable set is partitioned as $Y = Y_1\sqcup
Y_2\sqcup\cdots \sqcup Y_d$ where for each $j\in [d]$, $Y_j =
\{y_{ij}\}_{i=1}^n$. An ABP $B$ is homogeneous set-multilinear if each edge in the $j^{th}$ layer of the ABP is labelled by linear forms over $Y_j$. For ROABP, a different variable is used for each layer, and the edge labels are univariate polynomials. Therefore, an ROABP of $d$ layers can be represented as $ \boldsymbol{c} \cdot M_1(v_1) M_2(v_2) \cdots M_{v_d}(d) \cdot\boldsymbol{b}$. We say that the ROABP respects the variable order $v_1 < v_2 < \cdots < v_d$.


\subsubsection*{Identity testing results}

For the black-box case, Forbes and Shpilka \cite{FS13}, have shown an
efficient construction of quasipolynomial-size hitting set for
noncommutative ABPs. Consider the class of noncommutative ABPs of
width $w$, and depth $d$ computing polynomials in $\F\angle{X}$. The
result of Forbes-Shpilka provide an explicit construction (in
quasipolynomial-time) of a set $\mathcal{H}_{w,d,n}$ contained in
$\mathbb{M}_{d+1}(\F)$, such that for any ABP (with parameters $w$ and
$d$) computing a nonzero polynomial $f$, there always exists
$\ubar{u}\in\mathcal{H}_{w,d,n}$ such that
$f(\ubar{u})\neq 0$.

\begin{theorem}[Forbes-Shpilka \cite{FS13}]\label{forbesshpilka}
For all $w,d,n \in \mathbb{N}$, if $|\F|\geq \poly(d,n,w)$, then there is a
hitting set $\mathcal{H}_{w,d,n} \subset \M_{d+1}(\F)$ for
noncommutative \text{ABP}s of parameters $w,d,n$ such that
$|\mathcal{H}_{w,d,n}\mid \leq (wdn)^{O(\log d)}$ and there is a
deterministic algorithm to output the set $\mathcal{H}_{w,d,n}$ in
time $(wdn)^{O(\log d)}$.
\end{theorem}

Next, we define the concept of \emph{strong} hitting set. 
\begin{definition}
For a class of rational functions (resp. polynomials) a hitting set $\H$ is strong if any nonzero rational function (resp. polynomial) in that class evaluates to an invertible matrix at some point in $\H$.  
\end{definition}

In our proof, we also need the hitting set for ROABPs of unknown order \cite{AGKS15}. 
\begin{theorem}\cite{AGKS15}\label{thm:roabpanyorder}
Given the parameters $n, w, \delta$, in deterministic quasipolynomial-time, one can construct a hitting set $\H$ of size $(n w \delta)^{O(\log n)}$ for $n$-variate ROABPs (unknown order) of width $w$ and the degree of each variable is bounded by $\delta$.   
\end{theorem}




\subsubsection*{Recognizable series}
A comprehensive treatment is in the book by Berstel and Reutenauer~\cite{BR11}. We will require the following concepts. Recall that $\F\dangle{\ubar{x}}$ is the formal power series ring over a field $\F$. A series $S$ in $\F\dangle{\ubar{x}}$ is \emph{recognizable} if it has the following linear representation: for some integer $s$, there exists a row vector $\ubar{c}\in \F^{1\times s}$, a column vector $\ubar{b}\in \F^{s\times 1}$ and an $s\times s$ matrix $M$ whose entries are homogeneous linear forms over $x_1, \ldots, x_n$ i.e. $\sum_{i=1}^n \alpha_ix_i$ such that $S = \ubar{c}\left(\sum_{k\geq 0}M^k\right)\ubar{b}$. Equivalently, $S = \ubar{c}(I-M)^{-1}\ubar{b}$. We say, $S$ has a representation $(\ubar{c}, M, \ubar{b})$ of size $s$.


 
The following theorem is a basic result in algebraic automata theory.
\begin{theorem}\label{thm:sch-finite}
A recognizable series with representation $(\ubar{c}, M, \ubar{b})$ of size $s$ is nonzero if and only if $\ubar{c}\left(\sum_{k\leq s-1}M^k\right)\ubar{b}$ is nonzero.
\end{theorem}

It has a simple linear algebraic proof~\cite[Corollary 8.3, Page 145 ]{Eilenberg74}. This result is generally attributed to Sch\"{u}tzenberger. For the purpose of this paper, the theorem is used to apply that the truncated series is computable by a small noncommutative ABP therefore reducing zero-testing of recognizable series to the identity testing of noncommutative ABPs.  
  

\subsection{Matrix Coefficient Realization Theory} 
For a detailed exposition of this theory, see the work of Vol\v{c}i\v{c}~\cite{vol18}. Recall that, $\M_m(\F)$ is the $m\times m$ matrix algebra over $\F$. A \emph{generalized word} or a \emph{generalized monomial} in $x_1,\ldots, x_n$ over $\M_m(\F)$ allows the matrices to interleave between variables. More formally, a generalized word over $\M_m(\F)$ is of the following form: $a_0 x_{k_1}a_2\cdots a_{d-1}x_{k_d}a_{d}$ where $a_i\in \M_m(\F)$. A generalized polynomial over $\M_m(\F)$ is obtained by a finite sum of generalized monomials in the ring $\M_m(\F)\angle{\ubar{x}}$. Similarly, a generalized series over $\M_m(\F)$ is obtained by infinite sum of generalized monomials in the ring $\M_m(\F)\dangle{\ubar{x}}$.

A generalized series (resp. polynomial) $S$ over $\M_m(\F)$ admits the following canonical description. Let $E=\{e_{i,j}, 1\leq i,j\leq m\}$ be the set of matrix units. Express each coefficient matrix $a$ in $S$ in the $E$ basis by a $\F$-linear combination and then expand $S$. Naturally each monomial of degree-$d$ in the expansion looks like $e_{i_0,j_0} x_{k_1} e_{i_1,j_1} x_{k_2} \cdots e_{i_{d-1},j_{d-1}} x_{k_d} e_{i_d,j_d}$ where $e_{i_l,j_l}\in E$ and $x_{k_{l}}\in \ubar{x}$. We say the series $S$ (resp. polynomial) is identically zero if and only if it is zero under such expansion i.e. the coefficient associated with each generalized monomial is zero.   

The evaluation of a generalized series over $\M_m(\F)$ is defined on any $k'm\times k'm$ matrix algebra for some integer $k'\geq 1$ \cite{vol18}. To match the dimension of the coefficient matrices with the matrix substitution, we use an inclusion map $\iota: \M_m(\F)\to \M_{k'm}(\F)$, for example, $\iota$ can be defined as $\iota(a) = a\otimes I_{k'}$ or $\iota(a) = I_{k'}\otimes a$. We now define the evaluation of a generalized series (resp. polynomial) over $\M_m(\F)$ in the following way. Any degree-$d$ generalized word  $a_0x_{k_1}a_1\cdots a_{d-1}x_{k_d}a_{d}$ over $\M_m(\F)$ on a matrix substitution $(p_1,\ldots, p_n)\in \M^n_{k'm}(\F)$ evaluates to $$ \iota(a_0) p_{k_1} \iota(a_1)\cdots \iota(a_{d-1}) p_{k_d} \iota(a_d) $$under some inclusion map $\iota:\M_m(\F)\to \M_{k'm}(\F)$.  In ring theory, all such inclusions are known to be compatible by the Skolem-Noether theorem~\cite[Theorem 3.1.2]{row80}. Therefore, if a series $S$ is zero with respect to some inclusion map $\iota: \M_m(\F)\to \M_{k'm}(\F)$, then it must be zero w.r.t. any such inclusions. The equivalence of the two notions of zeroness follows from the proof of \cite[Proposition 3.13]{vol18}. 

We now recall the definition of a recognizable generalized series from the same paper.
\begin{definition}
A generalized series $S$ in $\M_m(\F)\dangle{\ubar{x}}$ is \emph{recognizable} if it has the following linear representation. For some integer $s$, there exists a row-tuple of matrices $\boldsymbol{c}\in (\M_m(\F))^{1\times s}$, and $\boldsymbol{b}\in (\M_m(\F))^{s\times 1}$ and an $s\times s$ matrix $M$ whose entries are homogeneous generalized linear forms over $x_1, \ldots, x_n$ i.e. $\sum_{i=1}^n p_ix_iq_i$ where each $p_i,q_i\in \M_m(\F)$ such that 
$S = \boldsymbol{c}(I-M)^{-1}\boldsymbol{b}$. We say, $S$ has a linear representation $(\boldsymbol{c}, M, \boldsymbol{b})$ of size $s$ over $\M_m(\F)$.
\end{definition} 

In \cite{vol18}, Vol\v{c}i\v{c} shows the following result.

\begin{theorem}~\cite[Corollary~5.1, Proposition 3.13]{vol18}\label{rit-connection}
Given a noncommutative rational formula $\r$ of size $s$ over $x_1, \ldots, x_n$ and a matrix tuple $\ubar{p}\in \M^n_m(\F)$ in the domain of definition  of $\r$, $\r(\ubar{x} + \ubar{p})$ is a recognizable generalized series with a representation of size at most $2s$ over $\M_m(\F)$. Additionally, $\r(\ubar{x})$ is zero in the free skew field if and only if $\r(\ubar{x}+\ubar{p})$ is zero as a generalized series. 
\end{theorem}

\begin{proof}
For the first part, see Corollary 5.1 and Remark 5.2 of \cite{vol18}. 

To see the second part, let $\r(\ubar{x})$ is zero in the free skew field. Then the fact that $\r(\ubar{x} + \ubar{p})$ is a zero series follows from Proposition 3.13 of \cite{vol18}. 
If $\r(\ubar x)$ is nonzero in the free skew field, then there exists a matrix tuple $(q_1, \ldots, q_n)\in \M^n_l(\F)$ such that $\r(\ubar q)$ is nonzero. W.l.o.g. we can assume $l = k'm$ for some integer $k'$. Fix an inclusion map $\iota: \M_m(\F)\to \M_{k'm}(\F)$. Define a matrix tuple $(q'_1, \ldots, q'_n)\in \M^n_{k'm}(\F)$ such that $q'_i = q_i - \iota(p_i)$. Therefore, the series $\r(\ubar x +\ubar p)$ on $(q'_1, \ldots, q'_n)$ evaluates to $\r(\ubar{q})$ under the inclusion map $\iota$, hence nonzero~\cite[Remark~5.2]{vol18}. Therefore, $\r(\ubar x + \ubar p)$ is also nonzero.
\end{proof}
\begin{remark}\label{remark:explicit-point}
More explicitly we can say the following which is already outlined in \cite[Section 5]{vol18}. For  inclusion map $\iota :\M_m(\F)\to \M_{k'm}(\F)$ 
\[
\r(\ubar{q} + \iota(\ubar{p})) = \iota(\boldsymbol{c}) \left(I_{2sk'm} - \sum_{j=1}^n \iota(A^{x_j})(\ubar{q})\right)^{-1}\iota(\boldsymbol{b}).
\]
\end{remark}








\subsection{Cyclic Division Algebra}\label{sec:cyclic}

We briefly recall cyclic division algebras and their construction
\cite[Chapter 5]{Lam01}. Let $F=\Q(z)$, where $z$ is a commuting
indeterminate. Let $\omega$ be an $\ell^{th}$ primitive root of unity. To
be specific, let $\omega= e^{2\pi i/\ell}$. Let
$K=F(\omega)=\Q(\omega,z)$ be the cyclic Galois extension of $F$ obtained by
adjoining $\omega$. The elements of $K$ are polynomials in $\omega$ (of
degree at most $\ell-1$) with coefficients from $F$.

Define $\sigma:K\to K$ by letting $\sigma(\omega)=\omega^k$ for some $k$
relatively prime to $\ell$ and stipulating that $\sigma(a)=a$ for all
$a\in F$. Then $\sigma$ is an automorphism of $K$ with $F$ as fixed
field and it generates the Galois group $\Gal(K/F)$.

The division algebra $D=(K/F,\sigma,z)$ is defined using a new
indeterminate $x$ as the $\ell$-dimensional vector space:
\[
D = K\oplus Kx\oplus \cdots \oplus Kx^{\ell-1},
\]
where the (noncommutative) multiplication for $D$ is defined by
$x^\ell = z$ and $xb = \sigma(b)x$ for all $b\in K$. That $D$ is a
division algebra of dimension $\ell^2$ over $F$ is well known
\cite[Theorem 14.9]{Lam01}. Its elements have matrix representations in
$K^{\ell \times \ell}$ (the regular matrix representation defined by
multiplication from the left) given below:

The matrix representation $M(x)$ of $x$ is:

\[
        M(x)[i,j] = \begin{dcases}
                        1 & \text{ if } j=i+1, i\le \ell-1 \\
                        z & \text{ if } i=\ell, j=1\\
                        0 & \text{ otherwise.}
                    \end{dcases}
\]

$$
M(x)=\begin{bmatrix}
    0       & 1 & 0 & \cdots & 0 \\
     0       & 0 & 1 &\cdots  &  0 \\
      \vdots & \vdots &\ddots &\ddots  &  \vdots \\
      0       & 0 &\cdots                  & 0 &1 \\
      z       & 0 & \cdots & 0 & 0
\end{bmatrix}.
$$

For each $b\in K$ its matrix representation $M(b)$ is:

\[
        M(b)[i,j] = \begin{dcases}
                        b & \text{ if } i=j=1 \\
                        \sigma^{i-1}(b) & \text{ if } i=j, i\ge 2\\
                        0 & \text{ otherwise.}
                    \end{dcases}
\]

\[M(b) = 
\begin{bmatrix}
b & 0 & 0 & 0 & 0 & 0  \\
0 & \sigma(b) & 0 & 0 & 0 & 0 \\
0 & 0 & \sigma^2(b) & 0 & 0 & 0 \\
0 & 0 & 0 & \ddots & 0 & 0 \\
0 & 0 & 0 & 0 & \sigma^{\ell-2}(b) & 0 \\
0 & 0 & 0 & 0 & 0 & \sigma^{\ell-1}(b)
\end{bmatrix}
\]
        
\begin{remark}
We note that $M(x)$ has a ``circulant'' matrix structure and $M(b)$ is
a diagonal matrix. For a vector $v\in K^\ell$, it is convenient to
write $\cir(v_1,v_2,\ldots,v_\ell)$ for the $\ell\times \ell$ matrix
with $(i,i+1)^{th}$ entry $v_i$ for $i\le \ell-1$, $(\ell,1)^{th}$
entry as $v_\ell$ and remaining entries zero. Thus, we have
$M(x)=\cir(1,1,\ldots,1,z)$.  Similarly, we write
$\diag(v_1,v_2,\ldots,v_\ell)$ for the diagonal matrix with entries
$v_i$.
\end{remark}

\begin{fact}
  The $F$-algebra generated by $M(x)$ and $M(b), b\in K$ is an
  isomorphic copy of the cyclic division algebra in the matrix algebra
  $\M_{\ell}(K)$.
\end{fact}

\begin{proposition}\label{circ-in-D}
  For all $b\in K$, $\cir(b,\sigma(b),\ldots,z\sigma^{\ell-1}(b)) = M(b)\cdot M(x)$.
\end{proposition}

Define $C_{i,j}= M(\omega^{j-1}) \cdot M(x^{i-1})$ for $1\leq i,j\leq \ell$. Observe that, $\B=\{C_{ij}, i,j \in [\ell]\}$ be a $F$-generating set for the division algebra $D$.

A standard fact is the following. 

\begin{proposition}\cite[Section 14(14.13)]{Lam01}\label{full-space}
Then $K$ linear span of $\B$ is the entire matrix algebra 
$\M_{\ell}(K)$. 
\end{proposition}



\section{Embedding Forbes-Shpilka Hitting Set inside a Division Algebra}\label{sec:embedding}

Given any noncommutative algebraic branching program of size $s$ computing a polynomial $h\in \F\angle{x_1,\ldots,x_n}$ of degree $\tilde{d}$, the hitting set $\H$ contains a matrix tuple $(p_1,\ldots,p_n)$ such that $h(p_1,\ldots,p_n)$ is nonzero. Forbes and Shpilka \cite{FS13} have shown a quasipolynomial-size hitting set construction contained in $\M^n_{\tilde{d}+1}(\F)$. For ABPs over $\Q$, we will show the construction of a hitting set $\H$ which is contained in $D^n$ such that $D$ is a cyclic division algebra of index $\ell$ where $\ell$ is suitably chosen depending on $n,\tilde{d}$ and $s$.


Before we present our construction, let us first recall the matrix substitutions from Forbes-Shpilka hitting set construction. The idea was to reduce the PIT of a noncommutative ABP to PIT of a commutative read-once oblivious ABP (ROABP) and to design a hitting set generator for the latter. Recall that, without loss of generality, we can assume that the given ABP computing $f$ in $\F\angle{x_1, \ldots, x_n}$ of degree $\tilde{d} = 2^d$ is an entry of the $\tilde{d}$-product of $r\times r$ matrices
$
M = A_1\cdot A_2\cdots A_{\tilde{d}},
$ 
 where the entries of each $A_i$ are homogeneous linear forms in $x_1,x_2,\ldots,x_n$.  The $j^{th}$ layer of this iterated product has the form $\sum_{i=1}^nA_{ij}x_i$, $1\le j\le \tilde{d}$, where $A_{ij}\in \F^{r\times r}$. The entries $M_{ij}$ of the matrix $M$ are homogeneous polynomials in $\F\angle{\ubar{x}}$. 
The polynomial $f$ is computed at some entry of $M$ as the output polynomial. 
 In~\cite{FS13}, they considered the following matrix substitution for each $x_i$:

\[
 M(x_i)=
\begin{bmatrix}
     0       & u_1^i & 0 & \cdots & 0 \\
     0       & 0 & u_2^i &\cdots  &  0 \\
     \vdots & \vdots &\ddots &\ddots  &  \vdots \\
     0       & 0 &\cdots                  & 0 &u_{\tilde{d}}^i \\
     0    & 0 & \cdots & 0 & 0
\end{bmatrix} .
\]
Evaluating the ABP $f$ on these matrices outputs a matrix whose $(1, \tilde{d}+1)^{th}$ entry is an ROABP.
The output matrix is the following product: 
\[
A_1(u_1)\cdot A_2(u_2)\cdots A_{{\tilde{d}}}(u_{\tilde{d}}).
\]
The PIT algorithm then follows from the construction of a hitting set generator for commutative ROABPs:
\[
\G_d : (\alpha_1, \alpha_2, \ldots, \alpha_{d}, \alpha_{d+1}) \mapsto (f_0({\alpha_1,\ldots, \alpha_{d}}, \alpha_{d+1}),f_1(\alpha_1,\ldots, \alpha_{d}, \alpha_{d+1}),\ldots,f_{2^d - 1}({\alpha_1, \ldots, \alpha_{d}}, \alpha_{d+1})),
\]
where each $f_i$ is a polynomial of degree $\poly(2^d,r,n)$. The actual points of the hitting set are obtained by choosing values for each variable $\alpha_i$ from a subset of scalars $U\subseteq \F$ of $\poly(2^d,r,n)$ size. This makes the size of the hitting set quasipolynomial. The final substitution for each $x_i$ variable in the noncommutative ABP is the following:

\begin{equation}\label{eq-FS13}
 M(x_i)=
\begin{bmatrix}
     0       & f^i_0(\alpha_1, \ldots, \alpha_{d+1}) & 0 & \cdots & 0 \\
     0       & 0 & f^i_1(\alpha_1, \ldots, \alpha_{d+1}) &\cdots  &  0 \\
     \vdots & \vdots &\ddots &\ddots  &  \vdots \\
     0       & 0 &\cdots                  & 0 &f^i_{2^d - 1}(\alpha_1, \ldots, \alpha_{d+1}) \\
     0    & 0 & \cdots & 0 & 0
\end{bmatrix}.
\end{equation}

Therefore, one approach to embed the matrix substitutions inside a cyclic division algebra $D=(K/F,\sigma,z)$ (where $F = \Q(z)$) of index $\ell$ (where $\ell$ is the index of $D$ which is larger than $2^d$ that we fix later) would be to find a hitting set generator
\[
\G_d : (\alpha_1, \alpha_2, \ldots, \alpha_{d}, \alpha_{d+1}) \mapsto (f_0({\alpha_1,\ldots, \alpha_{d}}, \alpha_{d+1}),f_2(\alpha_1,\ldots, \alpha_{d}, \alpha_{d+1}),\ldots,f_{2^d - 1}({\alpha_1, \ldots, \alpha_{d}}, \alpha_{d+1})),
\] 
with the following additional property: $f_{i+1}(\alpha_1, \ldots, \alpha_{d+1}) = \sigma(f_i(\alpha_1, \ldots, \alpha_{d+1}))$ for each $0\leq i\leq \ell - 2$. 
In that case, consider the following $\ell\times \ell$ matrix substitutions:

\[
 M(x_i)=\left[
\begin{array}{c c c c c | c c c}
     0       & f^i_0(\ubar{\alpha}) & 0 & \cdots & 0 &0 &\cdots  &0\\     
     0       & 0 & f^i_1(\ubar{\alpha}) & \cdots & 0 &0 &\cdots  &0\\
     \vdots & \vdots &\ddots &\ddots  &  \vdots &\vdots &\ddots &\vdots\\
     0       & 0 & 0 & \cdots & f^i_{\tilde{d} - 1}(\ubar{\alpha}) &0 &\cdots  &0\\
     0       & 0 & 0 & \cdots & 0 &f^i_{\tilde{d}}(\ubar{\alpha}) &\cdots  &0\\
     \hline
     \vdots & \vdots &\ddots &\ddots  &  \vdots &\vdots &\ddots &\vdots\\
     0       & 0 & 0 & \cdots & 0 &0 &\cdots  &f^i_{\ell-2}(\ubar{\alpha})\\     
     zf^i_{\ell - 1}(\ubar{\alpha}) &0 & 0 & \cdots & 0 &0 &\cdots  &0
\end{array}\right].
\]

Notice that the top-left $(\tilde{d}+1)\times (\tilde{d}+1)$ submatrix of this substitution is exactly the substitution described in Equation~\ref{eq-FS13}. Therefore, evaluating a degree-$\tilde{d}$ noncommutative ABP $B$ over $\{x_1, \ldots, x_n\}$ on these matrices will output the evaluation of corresponding ROABP in the $(1,\tilde{d}+1)^{th}$ entry as~\cite{FS13}. Moreover, by Proposition~\ref{circ-in-D}, we can ensure that each $M(x_i)$ is inside the cyclic division algebra $D$ assuming that each $f_i(\ubar{\alpha})\in K$. Therefore, the output will also be inside the division algebra $D$ only. To conclude, for a nonzero noncommutative ABP, the image will be nonzero and inside a division algebra, hence invertible.

Our goal is now to find a cyclic division algebra $D=(K/F,\sigma,z)$ (where $F = \Q(z)$) of index $\ell$ (more than $\tilde{d}$) and to construct a hitting set generator $\G_d : \ubar{\alpha} \mapsto (f_0(\ubar{\alpha}), \ldots, f_{2^d-1}(\ubar{\alpha}))$ for commutative ROABPs with the additional property that $f_{i+1}(\alpha_1, \ldots, \alpha_{d+1}) = \sigma(f_i(\alpha_1, \ldots, \alpha_{d+1}))$ for each $0\leq i\leq \ell-2$.

We now examine the Forbes-Shpilka construction to incorporate these
aspects. The construction is recursive. 
Suppose that we have the
construction for degree $2^{d-1}$.

The hitting set for degree $2^{d}$ is obtained in \cite{FS13} by
combining two copies of the hitting set for degree $2^{d-1}$ using the
following key technical lemma \cite[Lemma 3.7]{FS13}, rephrased below
in somewhat different notation.

Let $p_{\ell'}(v), 1\le \ell'\le r^2$ denote the Lagrangian interpolation
polynomials, interpolating values from $[r^2]$. Each $p_{\ell'}(v)$ is
univariate with integer coefficients of degree less than $r^2$.


\begin{lemma}\label{fs13-lemma}{\rm\cite[Lemma~3.7]{FS13}}
  Let $M_i$ and $N_i, 0\le i\le 2^{d-1}-1$, be $r\times r$ matrices with
  entries from $\F[x]$ of degree less than $n$. Let
  $(f_0(u),f_1(u),\ldots,f_{2^{d-1}-1}(u))\in \F[u]$ be polynomials of degree
  at most $m$. Let $\omega \in\F$ (or in an extension field) be an element of order at
  least $(2^dnm)^2$. Define polynomials in one indeterminate $v$:
  \begin{eqnarray*}
    f'_i & = & \sum_{\ell'=1}^{r^2} f_i(\omega^{\ell'} \alpha_d)p_{\ell'}(v),~ 0\le i\le 2^{d-1}-1\\
    f'_{i+2^{d-1}} & = & \sum_{\ell'=1}^{r^2} f_i((\omega^{\ell'} \alpha_d)^\mu)p_{\ell'}(v),~ 0\le i\le 2^{d-1}-1,
  \end{eqnarray*}
  where $\mu = 2^{\kappa+d-1}+1$ and $\kappa$ is chosen such that $2^\kappa\ge 2^d nm$.

 Then, for all but at most $(2^dnmr)^2$ many values of $\alpha_d$, the
 $\F$-linear span of the matrix coefficients of the matrix product
 $\prod_{i=0}^{2^{d-1}-1}M_i(f_i(x))\prod_{i=0}^{2^{d-1}-1}N_i(f_i(y))$
 is contained in the $\F$-linear span of the matrix coefficients of the
 product
 $\prod_{i=0}^{2^{d-1}-1}M_i(f'_i(v))\prod_{i=2^{d-1}}^{2^d-1}N_i(f'_i(v))$.
\end{lemma}  

Lemma~\ref{fs13-lemma} essentially gives the construction for going
from the degree $2^{d-1}$ hitting set generator to the degree $2^d$ hitting set generator
as proved in \cite{FS13}.

\begin{remark}
In our modified construction we will use
different roots of unity (for the element $\omega$) for different stages of the recursive
construction. In particular, roots of unity $\omega_i, i<d$, used in
stages $i<d$ will be of lower order.  We explain below in detail, the
choice of the parameters: $\ell$, $\kappa$, $\omega_i$ and $\alpha_i$ for
the modified construction.
\end{remark}

We now \emph{adapt} Lemma~\ref{fs13-lemma} to ensure the additional
properties that will guarantee that the points of the hitting set are
from $D^n$, for a suitably large cyclic division algebra $D$.


Let $\ell$ be the index of $D$. We set $\ell=2^L$, where $L$ is to be determined below. Thus,
$\omega=e^{\frac{2\pi}{2^L}}$ is a $2^L$-th primitive root of
unity. Let $F=\Q(z)$ and $K=F(\omega,z)$ which gives the cyclic
division algebra $D=(K/F,\sigma,z)$ where we fix the $K$-automorphism
$\sigma$ as
\[
\sigma(\omega)=\omega^{2^{\kappa}+1},
\]
and $\kappa$ will be suitably chosen in the following analysis, fulfilling
the constraints of Lemma~\ref{fs13-lemma} and some additional
requirements.

Let $\omega_i =\omega^{2^{a_i}}$ for $a_1>a_2>\cdots > a_d > 0$,
where $a_i$ are positive integers to be chosen. Let $K_i=F(\omega_i)$
be the cyclic Galois extension for $1\le i\le d$. This gives a tower
of extension fields
\[
F\subset F(\omega_1) \subset F(\omega_2)\subset \cdots \subset F(\omega_d) \subset F(\omega).
\]

We require two properties of $\omega_i, 1\le i\le d$.
\begin{enumerate}
\item For the hitting set generator $\G_i$ we will choose the root of
  unity as $\omega_i$ and the variable $\alpha_i$ will take values
  only in the set $W_i = \{\omega_i^j\mid 1\le j\le 2^{L-a_i}\}$.
\item We require that the $K$-automorphism $\sigma$ has the property
  that for all $1\le i\le d$ the map $\sigma^{2^i}$ fixes
  $\omega_i$. In fact we will ensure that $\sigma^{2^i}$ has
  $F(\omega_i)$ as its fixed field.
\end{enumerate}

We take up the second property. As $\sigma(\omega)=\omega^{2^{\kappa}+1}$,
we have $\sigma(\omega_i) = \omega^{2^{a_i}(2^{\kappa}+1)}$. Therefore
\[
\sigma^{2^i}(\omega_i) = \omega^{2^{a_i}(2^{\kappa}+1)^{2^i}}.
\]

Now, $(2^{\kappa}+1)^{2^i}=\sum_{j=0}^{2^i}{{2^i}\choose j} 2^{\kappa j}$. Choosing
$\kappa=L/2$, we have $\omega^{2^{\kappa j}}=1$ for $j\ge 2$. Therefore,
\[
\sigma^{2^i}(\omega_i) = \omega^{2^{a_i}(2^{i+\kappa}+1)} =
  \omega_i\cdot \omega^{2^{a_i+i+\kappa}}.
\]

We can set $a_i+i+\kappa = L$ for $1\leq i\le d$ to ensure that $\sigma^{2^i}$
fixes $\omega_i$. Putting $L=2\kappa$, we obtain
\begin{equation}\label{ai}
a_i = \kappa - i \textrm{ for } 1\le i\le d.
\end{equation}

It remains to choose $\kappa$. In the construction of our hitting set
generator $\G_i$, the parameter $\alpha_i$ will take values only in
$W_i$ defined above. We note that $|W_i|=2^{L-a_i}=2^{\kappa+i}$.  By
Lemma~\ref{fs13-lemma} there are at most $(2^d nmr)^2$ many bad values of
$\alpha_i$ for any $i$. Thus, it suffices to choose $\kappa$ such that $2^{\kappa}
> (2^d nmr)^2$. It suffices to set
\[
\kappa = 2d + \lceil 2\log_2 (nmr)\rceil +1.
\]

The choice of $\kappa$ determines the value of parameter $\mu$ in
Lemma~\ref{fs13-lemma}.

Coming back to the modified construction of $\G_d$, 
inductively, we can assume that the hitting set generator $\G_{d-1}: (\alpha_1, \ldots, \alpha_{d-1}, u)\mapsto (f_0(u),f_1(u),\ldots,f_{2^{d-1}-1}(u))$ (where for $0\leq i\leq 2^{d-1}-1$, the polynomial $f_i(u)\in K_{d-1}[u]$) has that property. Namely,
suppose $f_{i+1}(u)=\sigma(f_i(u))$ holds for all
$i\leq 2^{d-1}-2$. Now define $\G_d$ using
Lemma~\ref{fs13-lemma}. Since $p_{\ell'}(v)$ has only integer
coefficients, $\sigma(p_{\ell'}(v))=p_{\ell'}(v)$. Therefore, for $0\le
i\leq 2^{d-1}-2$ and for $2^{d-1}\le i \leq 2^d-2$ we have
$f'_{i+1}(v)=\sigma(f'_i(v))$.

Now, consider $i=2^{d-1}-1$. We need to ensure that
$\sigma(f'_{2^{d-1}-1}(v))=f'_{2^{d-1}}(v)$. Equivalently, we need to
    ensure that 
\[
    \sigma\left(\sum_{\ell'=1}^{r^2} f_{2^{d-1}-1}(\omega_d^{\ell'}
    \alpha_d)p_{\ell'}(v)\right) = \sum_{\ell'=1}^{r^2}
    f_1((\omega_d^{\ell'} \alpha_d)^\mu)p_{\ell'}(v).
\]

This is enforced by requiring that 

\[
    \sigma^{2^{d-1}}\left(\sum_{\ell'=1}^{r^2} f_1(\omega_d^{\ell'}
    \alpha_d)p_{\ell'}(v)\right) = \sum_{\ell'=1}^{r^2} f_1((\omega_d^{\ell'}
    \alpha_d)^\mu)p_{\ell'}(v).
\]

Since $\alpha_d$ will be chosen from $W_d$ (all powers of $\omega_d$),
we can write $\omega_d^{\ell'}\alpha_d = \omega_d^j$ for some $j$. Now,
$\sigma^{2^{d-1}} f_1(\omega_d^j) = f_1(\sigma^{2^{d-1}}(\omega_d^j))$
as $\sigma^{2^{d-1}}$ fixes all coefficients of $f_1$ (because
$f_1(u)\in K_{d-1}[u]$). Now,
\[
\sigma^{2^{d-1}}(\omega_d^j)) = \omega_d^{j\cdot
  (2^\kappa+1)^{2^{d-1}}}=\omega_d^{j(1+2^{d-1+\kappa})}=(\omega_d^\ell\alpha_d)^\mu,
\]
which verifies the choice of $\mu$ in Lemma~\ref{fs13-lemma} is
$1+2^{d-1+\kappa}$.

As shown in \cite{FS13}, the parameter $v$ (whose place holder is $\alpha_{d+1}$ in the description of $\mathcal{G}_d$) should vary over a set of size $\poly(2^d,n,m,r)$.  This way we ensure that 
$f_{i+1}=\sigma(f_{i})$ for $0\leq i\leq 2^d-2$. Now define $f_{2^d+j}=\sigma(f_{2^d + j-1})$ for $0\leq j\leq \ell-2^d-1$. The fact that $\G_d$ is indeed a generator follows from the span preserving property and the proof is identical to the proof of \cite[Lemma~3.19]{FS13}.

We now summarize the above description in the following theorem. 

\begin{theorem}\label{thm:embedded-forbes-shpilka}
In deterministic quasipolynomial-time, we can construct a hitting set $\H$ of size $(nr\tilde{d})^{O(\log \tilde{d})}$ in $D^n$ for the class of noncommutative polynomials in $\Q\angle{x_1,\ldots, x_n}$ computed by ABPs of width at most $r$ with $\tilde{d}$ many layers where the index of the cyclic division algebra $D$, the parameter $\ell (> \tilde{d})$ is bounded by $\poly(r,n,\tilde{d})$. 
\end{theorem}

Note that $\H$ is a strong hitting set for any such noncommutative ABP. 




\section{Strong Hitting Set for Generalized ABPs over Division Algebra}
\label{sec:genabp-hs}
In this section, we first define the notion of a generalized ABP, ABPs over a division algebra, and then show the construction of a quasipolynomial-size strong hitting set for generalized ABPs over a division algebra such that any nonzero generalized ABP will evaluate to an invertible matrix on some point in the hitting set.

\begin{definition}
A \emph{generalized ABP} over the matrix algebra $\M_m(\F)$ is defined in the same way as a noncommutative ABP, except the fact that
the linear forms labeling the edges are of the form $\sum_{i=1}^n
a_ix_i b_i,$ where $a_i, b_i\in \M_m(\F)$. Clearly, such an ABP computes a generalized polynomial in the generalized polynomial ring $\M_m(\F)\angle{X}$, where the polynomial is defined as the sum of products of the linear forms along all $s$-to-$t$ paths of the ABP, where $s$ is the source node and $t$ is the sink node of the directed acyclic graph underlying the ABP.

If linear forms labeling the edges of the ABP are of the form $\sum_{i=1}^n
a_ix_i b_i, a_i, b_i\in D$ where $D$ is a division algebra, we say it is a \emph{generalized ABP over division algebra $D$}.
\end{definition}  


Let $D = (K/F, \sigma, z)$ (here $F = \Q(z)$) be a cyclic division algebra of index $\ell$ as defined in Section \ref{sec:cyclic}. Let $\mathfrak{B} = \{C_{ij}\}_{i,j\in [\ell]}$ be an $F$-basis of $D$ for $i,j\in [\ell]$. Informally, our idea is to reduce the problem of finding strong hitting set for generalized ABPs over division algebra to the hitting set construction of a product of commutative ROABPs. 

\begin{lemma}\label{ref-lemma-divalg-witness}
For any nonzero generalized ABP $B$ of degree $d$ over $D\angle{\ubar{x}}$, there exists a substitution for each $x_k$ of the following form:
\[
M(x_k)=\begin{bmatrix}
    0       &p_{k1} & 0 & \cdots & 0 \\
     0       & 0 &p_{k2} &\cdots  &  0 \\
      \vdots & \vdots &\ddots &\ddots  &  \vdots \\
      0       & 0 &\cdots                  & 0 &p_{k(d-1)} \\
      p_{kd}       & 0 & \cdots & 0 & 0
\end{bmatrix},
\]
such that for each $l\in [d]$, $p_{kl}$ is in $D$ and image of $B$ is invertible on that substitution under the inclusion map $a\mapsto I_d\otimes a$ where $a\in D$.
\end{lemma}

\begin{proof}
Let $\ell$ be the index of the division algebra $D$. We first prove that for any nonzero generalized ABP $B$ of degree $d$ over $D\angle{\ubar{x}}$, there exists a substitution for each $x_k$ of the following form:
\[
M(x_k)=\begin{bmatrix}
    0       &q_{k1} & 0 & \cdots & 0 \\
     0       & 0 &q_{k2} &\cdots  &  0 \\
      \vdots & \vdots &\ddots &\ddots  &  \vdots \\
      0       & 0 &\cdots                  & 0 &q_{k (d-1)} \\
      q_{kd}       & 0 & \cdots & 0 & 0
\end{bmatrix},
\]
such that for each $l\in [d]$, $q_{kl}$ is in $\M_{\ell}(K)$ and the image of $B$ is nonzero on that substitution with a block-diagonal structure. To evaluate $B$ on such matrix substitution the coefficients $a\in D$ (which have matrix representations in $\M_{\ell}(K)$) are fit to the correct dimension using the inclusion map $\iota':\M_{\ell}(K)\to \M_{d\ell}(K)$ where $\iota'(a) = I_d\otimes a$.

Let $\psi$ be the substitution map that replaces each variables $\{x_k\}_{k\in [n]}$ by an $\ell \times \ell$ matrix of noncommuting variables $\{z_{ijk}\}_{i,j\in [\ell], k\in [n]}$. One can naturally extend the definition of $\psi: \M_{\ell}(K)\angle{\ubar{x}}\to \M_{\ell}(K\angle{\ubar{z}})$ i.e. $\psi$ maps a generalized polynomial over matrix algebra $\M_{\ell}(K)$ to an $\ell \times \ell$ matrix of noncommutative polynomials in $K\angle{\ubar{z}}$. Indeed, the map $\psi$ is identity preserving (see~\cite[Equation~3.10]{vol18} for example).  

Introduce a new set of commuting variables $\widetilde{Z} = \{\tilde{z}_{ijkl}\}$ where $i,j\in [\ell]$, $k\in [n]$ and $l\in [d]$ and consider the following substitution for each $x_k$:

\[
\widetilde{Z}_{k}=
\begin{bmatrix}
    0       &\widetilde{Z}_{k1} & 0 & \cdots & 0 \\
     0       & 0 &\widetilde{Z}_{k2} &\cdots  &  0 \\
      \vdots & \vdots &\ddots &\ddots  &  \vdots \\
      0       & 0 &\cdots                  & 0 &\widetilde{Z}_{k(d-1)} \\
      \widetilde{Z}_{kd}       & 0 & \cdots & 0 & 0
\end{bmatrix},
\]
where $\widetilde{Z}_{kl} = (\tilde{z}_{ijkl})_{1\leq i,j\leq \ell}$. 
In effect the substitution of the $x_k$ variables by the matrices $\widetilde{Z}_k$ is just set-multilinearization of $\psi(B)$ position-wise and hence identity preserving.  

What is the effect of this substitution on a degree-$d$ generalized word? To understand that consider a generalized word 
$w=a_0x_{k_1}a_1 x_{k_2}\cdots a_{d-1} x_{k_d} a_{d}$ where each $a_i\in \M_{\ell}(K)$. Observe that $w(\widetilde{Z})$ is a diagonal matrix (using the inclusion map $\iota'$)
with $(i,i)^{th}$ entry $a_0\widetilde{Z}_{k_1 \pi_i(1)}a_1\cdots a_{d-1}\widetilde{Z}_{k_d \pi_i(d)}a_{d}$ where $\pi_i, 1\leq i\leq d$ is the cyclic permutations on $[d]$ such that $\pi_i(1) = i$, $\pi_i(2) = i+1$ and so on. For example, consider the case $d = 3$. For a generalized word $a_0 x_1 a_1 x_2 a_2 x_3 a_3$, the image will be the following product:
\[
\begin{bmatrix}
a_0\\
&a_0\\
&&a_0
\end{bmatrix}
\begin{bmatrix}
&\widetilde{Z}_{11}\\
&&\widetilde{Z}_{12}\\
\widetilde{Z}_{13}
\end{bmatrix}
\begin{bmatrix}
a_1\\
&a_1\\
&&a_1
\end{bmatrix}
\begin{bmatrix}
&\widetilde{Z}_{21}\\
&&\widetilde{Z}_{22}\\
\widetilde{Z}_{23}
\end{bmatrix}
\begin{bmatrix}
a_2\\
&a_2\\
&&a_2
\end{bmatrix}
\begin{bmatrix}
&\widetilde{Z}_{31}\\
&&\widetilde{Z}_{32}\\
\widetilde{Z}_{33}
\end{bmatrix}
\begin{bmatrix}
a_3\\
&a_3\\
&&a_3
\end{bmatrix},
\]
which outputs the following diagonal matrix:
\[
\begin{bmatrix}
a_0\widetilde{Z}_{11}a_1\widetilde{Z}_{22}a_2\widetilde{Z}_{33}a_3\\
&a_0\widetilde{Z}_{12}a_1\widetilde{Z}_{23}a_2\widetilde{Z}_{31}a_3\\
&&a_0\widetilde{Z}_{13}a_1\widetilde{Z}_{21}a_2\widetilde{Z}_{32}a_3
\end{bmatrix}.
\]

Let $B=\sum a_0 x_{k_1} a_1 x_{k_2} a_2 \ldots a_{d-1} x_{k_d} a_{i_{d}}$. 
So the $(i,i)^{th}$ entry of $B(\widetilde{Z})$ is 
\[
B^{\pi_i}= \sum a_0 \widetilde{Z}_{k_1 \pi_i(1)} a_1 \widetilde{Z}_{k_2 \pi_i(2)}  \ldots a_{d-1}\widetilde{Z}_{k_d \pi_i(d)} a_{i_{d}}. \]

Hence the final output matrix will be the following: 
\[
B(\widetilde{Z})=\begin{bmatrix}
B^{\pi_1}\\
&B^{\pi_2}\\
&&\ddots\\
&&&& B^{\pi_d}
\end{bmatrix}.
\]
We now claim the following.

\begin{claim}
For each $i\in [d]$, $B^{\pi_i}$ is nonzero.
\end{claim}

\begin{proof}
As $B$ in $D\angle{\ubar{x}}$ is nonzero and $\psi$ is an identity preserving substitution, $\psi(B)\in \M_{\ell}(K\angle{\ubar{z}})$ is also nonzero. We now consider the entry-wise set-multilinearization of $\psi(B)$ with respect to the cyclic permutation $\pi_i$ i.e. encoding any word using $\pi_i(j)$ as the position index for the $j^{th}$ position for each entry of $\psi(B)$. Notice that, it outputs the matrix $B^{\pi_i}$. Moreover, as $\psi(B)$ is nonzero, $B^{\pi_i}$ must be nonzero as set-multilinearization preserves identity.
\end{proof}

Hence, there exist  substitutions $q_{kl}$ from $\M_{\ell}(K)$ for the $\widetilde{Z}$ variables such that $B$ is nonzero. 


Now we use Fact \ref{full-space} which says that $K$-linear span of 
$\mathfrak{B}$ is the entire matrix algebra $\M_{\ell}(K)$. 
The above argument shows that if we replace each 
$q_{kl}$ in $M(x_k)$ by a linear combination $$\sum_{i,j} y_{ijkl} C_{ij},$$ each diagonal block matrix of the output matrix obtained from the image of $B$ on this evaluation is still nonzero over the $\{y_{ijkl}\}$ variables. 
We now find substitutions for the $Y$ variables from the ground field $F$ to make each diagonal block matrix nonzero. As any $F$-linear combination of $C_{ij}$ is inside the division algebra, each such linear combinations is in $D$. So, define $p_{kl}=\sum_{i,j} \beta_{ijkl} C_{ij}\in D$ where 
$\beta_{ijkl}$ are the substitutions for $y_{ijkl}$ variables from $F$. In fact the values for the variables $\beta_{ijkl}$ can be found from $\Q$ itself by a standard use of Polynomial Identity Lemma \cite{Zip79,Sch80}. 
Notice that, each diagonal block will also be inside $D$.
Since each diagonal block matrix is nonzero and inside $D$, hence invertible. Therefore, the image of $B$ is also invertible on the chosen matrix tuple. 
\end{proof}

We are now ready to prove the main result of this section.

\begin{theorem}\label{thm:gen-abp-hitting}
Given the parameters $n,\ell,r,d$, in deterministic quasipolynomial-time we can construct strong hitting set  $\H'$ of size $(nrd\ell)^{O(\log n d \ell)}$ for any nonzero generalized ABP $B$ of degree $d$ and width $r$ over $D\angle{\ubar{x}}$ where $\ell$ is the index of $D$. 
\end{theorem}
\begin{proof}
By Lemma \ref{ref-lemma-divalg-witness}, we know that there exists matrix tuple 
$(p_1, \ldots, p_n)$ in $\M^n_{d\ell}(K)$ 
of the following form 
\[
p_k=M(x_k)=\begin{bmatrix}
    0       &p_{k1} & 0 & \cdots & 0 \\
     0       & 0 &p_{k2} &\cdots  &  0 \\
      \vdots & \vdots &\ddots &\ddots  &  \vdots \\
      0       & 0 &\cdots                  & 0 &p_{k(d-1)} \\
      p_{kd}       & 0 & \cdots & 0 & 0
\end{bmatrix},
\]
where each $p_{k l}\in D :1\leq k\leq n, 1\leq l\leq d$ such that $B(p_1, p_2, \ldots, p_n)$ is an invertible matrix. 

Write each $p_{kl}$ as 
$p_{kl} = \sum_{i,j\in [\ell]} y_{ijkl}C_{ij}$ for some commuting indeterminates $Y = \{y_{ijkl}\}$ whose values we need to determine. On such a substitution, $B$ evaluates to the following matrix:  

\[
\begin{bmatrix}
B_1\\
&B_2\\
&&\ddots\\
&&&&B_d
\end{bmatrix}.
\]
where each $B_l, 1\leq l\leq d$ is nonzero by Lemma \ref{ref-lemma-divalg-witness} (using the inclusion map $\iota'$). We now observe the following.

\begin{claim}
For each $l\in [d]$, $B_l$ is a matrix of commutative set-multilinear ABPs each of width $r\ell$.
\end{claim}

\begin{proof}
To see this, consider the matrix $B_1$. We can think of $B_1$ as the matrix obtained by substituting $p_{kl}$ for $x_k$ in layer $l$ of the input generalized ABP $B$ over $D$ of index $\ell$. This computation can also be thought of by making $\ell$-many copies of each node in $B$. 

More precisely, each coefficient $a\in D$ in $B$ has a $\ell\times \ell$ matrix representation over $K$. Now consider each edge $\sum_{k=1}^n a_k x_k b_k$ between the layer $l$ and $l+1$. Since $x_k$ is replaced by $p_{kl}$ and $a_k, b_k \in D$, this edge can be replaced by an $\ell\times \ell$ bipartite graph such that for each $i,j\in [\ell]$, the edge connecting the $i^{th}$ node (from left) to the $j^{th}$ node (to right) is  labeled by the $(i,j)^{th}$ entry of the product of $a_k p_{kl} b_k$, a linear form over $K[Y]$. Clearly, it produces an $\ell$-input $\ell$-output setmultilinear ABP of width $r\ell$. Therefore, each entry in $B_1$ is computed by a set-multilinear ABP of width $r\ell$ and degree $d$. The situation for other $B_l : 2\leq l\leq d$ are similar. 
\end{proof}

Therefore we can use a hitting set generator for commutative set-multilinear ABPs of width $r\ell$ and degree $d$ to obtain a point such that the image for each $B_l$ is nonzero on that evaluation.
 
 However, our goal is to obtain an invertible image for the image of $B$. In other words, we want a substitution of $Y$ variables for which each $B_l$ would be invertible. Notice that, for some substitution of $Y$ variables from $F$, if at least one entry of $B_l$ is nonzero, then $B_l$ is also invertible as the image of $B_l$ would be inside $D$. Hence, to obtain a strong hitting set for the input generalized ABP over $D$ (equivalently, to obtain a substitution on which the product of the matrices $B_l, 1\leq l \leq d$ is invertible), it suffices to obtain a hitting set for the product of set-multilinear ABPs (product of one nonzero entries of each $B_l$). We do this by first converting each set-multilinear ABP to an ROABP encoding each $y_{ijkl}$ to $v^{(\ell+1)^2i+ (\ell+1) j+ k}_l$\footnote{Note that by the choice, $\ell$ is larger than $n$ and $d$.}. However, notice that, each encoded $B_l$ has a different variable partition on the ROABPs. More precisely, the ROABPs computed in the $(l,l)^{th}$ diagonal block follows the following partition: $$v_l < v_{l+1} < \ldots < v_d< v_1< \ldots < v_{l-1}.$$

We now use the hitting set generator for commutative ROABPs constructed in Theorem \ref{thm:roabpanyorder}. 
Moreover, as it is a generator, it also works for a product of ROABPs of different order. This is a standard argument using union bound that the choice for the seed of the generator should avoid a slightly larger set. The ROABPs are $n d\ell^2$-variate, $d$-degree, and of width $\ell r$. Thus by Theorem \ref{thm:roabpanyorder}, the size of the hitting set for them is $(n d \ell r)^{O(\log n d \ell)}$.     

Using this, we can now find a substitution for the $v_l$ variables such that each $B_l$ is invertible, hence $B$ is also invertible.

This gives us a hitting set $\mathcal{H}$ under the inclusion map $\iota': \M_{\ell}(K)\to \M_{d\ell}(K)$ where $\iota'(a) = I_d\otimes a$. 
However for the purpose of Section \ref{sec:all-together}, we find a hitting set $\mathcal{H}'$ under the inclusion map $\iota: \M_{\ell}(K)\to \M_{d\ell}(K)$ where $\iota(a) = a\otimes I_d$. Although it is technically possible to work with two inclusion maps thanks to Remark \ref{remark:explicit-point}, we find it mathematically nicer to work with a single inclusion map. 
For this we explicitly find a permutation matrix $q_0$ of dimension $d\ell$ such that $q_0 (I_d \otimes a) q^{-1}_0=a\otimes I_d$ for all $a\in \M_{\ell}(K)$. Once we find $q_0$, the final hitting set can be defined as 
$\mathcal{H'} = \{(q_0p_1q^{-1}_0, \ldots, q_0p_nq^{-1}_0)\mid \ubar{p}\in \mathcal{H}\}$. 
To see this, let 
\[
B = \sum a_0x_{k_1}a_1\cdots a_{d-1}x_{k_d}a_d.
\]
Let $M=B(q_1, \ldots, q_n)$ is an invertible matrix for $\ubar{q}\in \mathcal{H}$. 
We know that,
\begin{equation}\label{eq-skolem}
\sum (I_d\otimes a_0)q_{k_1}(I_d\otimes a_1)\cdots (I_d\otimes a_{d-1}) q_{k_d}(I_d\otimes a_d) = M.
\end{equation}
By conjugating $M$ with $q_0$, obtain the following: 
\begin{equation}\label{eq-skolem}
\sum q_0(I_d\otimes a_0) q^{-1}_0 q'_{k_1} q_0(I_d\otimes a_1)q^{-1}_0\cdots q_0(I_d\otimes a_{d-1})q^{-1}_0 q'_{k_d} q_0(I_d\otimes a_d)q^{-1}_0 = q_0 M q^{-1}_0
\end{equation}
where $q'_{k_j} = q_0 q_{k_j} q^{-1}_0$. In other words $B(q'_1, \ldots, q'_n)$ is the invertible matrix $M'=q_0 M q^{-1}_0$ under the inclusion map 
$\iota$. In the following, we show that the permutation matrix $q_0$ can be constructed explicitly.

\paragraph*{Explicit construction of $q_0$:}

Let us divide the $d\ell$ rows in group of $d$ rows 
as $1,\ldots, d, d+1, \ldots, 2d, \ldots, (\ell-1)d, \ldots, d\ell$. 
For the group of rows $id+1$ to $id+d$ (for $0\leq i\leq (\ell-1)$), set the $(id+j, (j-1)\ell+(i+1))^{th}$ entry for $1\leq j\leq d$ to be 1 and remaining entries to be zero. To elaborate it, we consider the case where $\ell=2$ and $d=3$ and give an illustrative example. In this case, $q_0$ and $q^{-1}_0$ are the following matrices.
\[
q_0 = \begin{bmatrix}
1\\
&&1\\
&&&&1\\
&1\\
&&&1\\
&&&&&1
\end{bmatrix},
\quad\quad
q^{-1}_0 = \begin{bmatrix}
1\\
&&&1\\
&1\\
&&&&1\\
&&1\\
&&&&&1
\end{bmatrix}.
\]
\[
\text{Let, }
a = 
\begin{bmatrix}
1 &2\\
3 &4
\end{bmatrix},
\quad\quad\text{then, }
I_3\otimes a = \left[\begin{array}{c c | c c | c c}
1&2\\
3&4\\
\hline
&&1&2\\
&&3&4\\
\hline
&&&&1&2\\
&&&&3&4
\end{array}
\right].
\]Consider the effect of $q_0$.
\[q_0aq^{-1}_0 = \left[\begin{array}{c c c | c c c}
1&&&2\\
&1&&&2\\
&&1&&&2\\
\hline
3&&&4\\
&3&&&4\\
&&3&&&4
\end{array}\right] = a\otimes I_3.
\]
\end{proof}

\section{Putting all together}\label{sec:all-together} 

In this section we prove our main result, construction of a hitting set for noncommutative rational formulas of inversion height two. An intermediate step is to construct a strong hitting set for rational formulas of inversion height one. En route to our proof, we crucially use the connection of rational identity testing with the identity testing of generalized ABPs. We make it explicit in Proposition \ref{prop-gen-Sch}. 
But before this, we note a basic result that we use throughout the section. 

\begin{lemma}\label{lem:bivariate-degree}
Let $\r\in \F\newbrak{\ubar{x}}$ be a rational formula of size $s$. Let $\ubar{p}=(p_1,\ldots,p_n)\in \M^n_m(\F(t_1, t_2))$ be an $n$-tuple of matrix of bivariate rational functions where the degrees of the numerator and denominator polynomials over $t_1, t_2$ at each entry are at most $d'$ and $\r$ is defined at $\ubar{p}$. Then, evaluating $\r$ on $\ubar{p}$ outputs $\r(\ubar{p})\in \M_m(\F(t_1, t_2))$ such that each entry of the output matrix is of form $\frac{P(t_1,t_2)}{Q(t_1, t_2)}$ where $P$ and $Q$ are bivariate polynomials of degree at most $O(smd')$.
\end{lemma}

\begin{proof}
As already stated in Section \ref{sec:intro} that $\r$ has a linear pencil $L$ of size (at most) $2s$ such that for any tuple $\ubar{p}$, $\r(\ubar{p})$ is defined if and only if $L(\ubar{p})$ is invertible ~\cite[Proposition~7.1]{HW15}. Moreover, $\r(\ubar{p}) = L^{-1}_{i,j}(\ubar{p})$ for some $(i,j)^{th}$ entry of $L$ i.e. $\r(\ubar{p})$ is the $(i,j)^{th}$ block of $L^{-1}(\ubar{p})$ thinking of it as a $2s\times 2s$ block matrix where each block is of size $m$. 
Notice that, if $L = \sum_{i=1}^n A_ix_i$, then $L(\ubar{p}) = \sum_{i=1}^n A_i\otimes p_i$. Therefore, $L(\ubar{p})$ is a $2sm\times 2sm$ matrix such that each entry is a polynomial over $t_1,t_2$ of degree at most $d'$. 
From the standard computation of matrix inverse, it is immediate that each entry of $L^{-1}(\ubar{p})$ (therefore, each entry of $\r(\ubar{p}))$ is a commutative rational function such that the numerator and the denominator are bivariate polynomials over $t_1, t_2$ with degree bound $O(smd')$. 
\end{proof}
Now we are ready to prove the main proposition. 

\begin{proposition}\label{prop-gen-Sch}
Let $\r$ be a noncommutative rational formula over $x_1,\ldots,x_n$ of size $s$ and $(q_1,\ldots,q_n)\in \M^n_m(\F)$ be a matrix tuple such that $\r$ is defined on $\ubar{q}$. Suppose, $\r(\ubar{x}+\ubar{q})$ is a recognizable generalized series over $\M_m(\F)\dangle{\ubar{x}}$ with a linear representation $(\boldsymbol{c}, M, \boldsymbol{b})$ of size at most $2s$ over $\M_m(\F)$. Define $S^{\{d\}} = \boldsymbol{c}\cdot M^d \cdot \boldsymbol{b}$ computing a generalized polynomial in $\M_m(\F)\angle{\ubar{x}}$. Then $\r$ is nonzero in $\F\newbrak{\ubar{x}}$ if and only if $S^{\{d\}}$ is nonzero for some $d\leq 2sm-1$. Additionally for sufficiently large $\F$,
\begin{enumerate}
\item For some matrix tuple $(p_1,\ldots,p_n)\in \M^n_{km}(\F)$, if $S^{\{d\}}$ is nonzero at $\ubar{p}$ for some $d\leq 2sm-1$ under the inclusion map $\iota: \M_m(\F)\to \M_{km}(\F)$ where $\iota(a) = a\otimes I_k$, then there exists an $\alpha \in \F$ such that $\r$ is nonzero at the following matrix tuple:
\[
(\alpha p_1 + q_1\otimes I_k, \ldots, \alpha p_n + q_n\otimes I_k).
\]


\item For some matrix tuple $(p_1,\ldots,p_n)\in \M^n_{km}(\F)$, if $S^{\{d\}}$ is invertible at $\ubar{p}$ for some $d\leq 2sm-1$ under the inclusion map $\iota: \M_m(\F)\to \M_{km}(\F)$ where $\iota(a) = a\otimes I_k$, then there exists an $\alpha \in \F$ such that $\r$ is invertible at the following matrix tuple:
\[
(\alpha p_1 + q_1\otimes I_k, \ldots, \alpha p_n + q_n\otimes I_k).
\]
\end{enumerate}
\end{proposition}

\begin{proof}
By Theorem \ref{rit-connection}, we know that $\r(\ubar{x})$ is zero if and only if $\r(\ubar{x} + \ubar{q})$ is zero. 
Let $Z=\{z_{i,j,k'}\}_{1\leq i,j\leq m, 1\leq k'\leq n}$ be a set of noncommuting variables. Consider a substitution map $\psi$ that substitutes each variable $x_{k'}, 1\leq k'\leq n$ of $\r(\ubar{x}+ \ubar{q})$ by an $m\times m$ matrix $Z_{k'}$ consisting of fresh noncommutative variables $\{z_{i,j,k'}\}_{1\leq i,j\leq m}$. Consider $\r(\psi(\ubar{x})+\ubar{q})$ and observe that, $\psi$ is an identity preserving and degree preserving substitution.

From the definition, $\r(\ubar{x}+\ubar{q})= \boldsymbol{c}(I-M)^{-1}\boldsymbol{b}$ where $M$ is of size at most $2s$ by Theorem \ref{rit-connection}. Therefore, $\r(\psi(\ubar{x})+\ubar{q}) = C(I-\psi(M))^{-1}B$, where 
it is convenient to think of $\boldsymbol{c}$ (respectively $\boldsymbol{b}$) as an $m\times 2ms$ (resp. $2ms\times m$) rectangular matrix $C$ (resp. $B$), and $\psi(M)$ as $2ms\times 2ms$ matrix. 

Observe that, for the matrix $\r(\psi(\ubar{x})+\ubar{q})$, the $(i,j)^{th}$ entry is the following recognizable series which has linear representation of size at most $2sm$: 
\[
\boldsymbol{C_i}(I - \psi(M))^{-1}\boldsymbol{B_j}
\]
where $\boldsymbol{C_i}$ is the $i^{th}$ row of $C$ and $\boldsymbol{B_j}$ is the $j^{th}$ column of $B$. 
If $\r(\ubar{x}+\ubar{q})$ is nonzero, then some $(i,j)^{th}$ entry of $\r(\psi(\ubar{x})+\ubar{q})$ is also nonzero. Clearly, the degree-$d$ truncated part of the matrix $\r(\psi(\ubar{x})+\ubar{q})$ is $\psi(S^{\{d\}})$. Moreover, for the matrix $\psi(S^{\{d\}})$, each entry is computed by a  noncommutative ABP of width $2sm$ and depth $d$ over $Z$ variables. By Theorem \ref{thm:sch-finite}, there exists a minimum $d\leq 2sm-1$ such that $\psi(S^{\{d\}})$ and thus $S^{\{d\}}$ is nonzero. Clearly $S^{\{d\}}$ is computable by a generalized ABP. 



\paragraph{Proof of part(1):} Now, for some matrix tuple $(p_1,\ldots,p_n)\in \M_{km}(\F)$, 
let $d\leq 2sm -1$ such that $S^{\{d\}}$ is nonzero at $\ubar{p}$ under the inclusion map $\iota : \M_m(\F)\rightarrow \M_{km}(\F)$ given by $\iota : a\rightarrow a\otimes I_k$. 
Consider the evaluation of $\r$ at $(tp_1 + q_1\otimes I_{k}, \ldots, tp_n + q_n\otimes I_{k})$ where $t$ is some commuting indeterminate. Let $M(t) = \r(tp_1 + q_1\otimes I_k, \ldots, tp_n + q_n\otimes I_k)$. We now interpret $M(t)$ in two ways. First, think of $M(t)$ as the evaluation of the generalized series $\r(\ubar{x}+\ubar{q})$ at $(tp_1, \ldots, tp_n)$ under the inclusion map $\iota : \M_m(\F)\rightarrow \M_{km}(\F)$ given by $\iota : a\rightarrow a\otimes I_k$. We can write $M(t) = t^d S^{\{d\}}(\ubar{p}) + M'(t)$ where $t$-degree of each term of the matrix $M'(t)$ is strictly more than $d$. Therefore, $M(t)$ is nonzero.

Another way to interpret $M(t)$ is to evaluate the rational formula $\r$ on $(t p_1 + q_1\otimes I_k, \ldots, t p_n + q_n\otimes I_k)$. Since $\r$ is a rational formula of size $s$, each entry of the matrix $M(t)$ is an element of the function field $\F(t)$. Moreover by Lemma \ref{lem:bivariate-degree}, the $t$-degrees of the numerator and denominator polynomials of each such commutative rational expression computed at all the nodes, are bounded by $\hat{d}=\poly(k s m)$. 
Therefore, the final choice of the parameter $t$ should be such that it avoids the zeros of the numerator and denominator polynomials involved in the computation of $M(t)$. This is clearly possible by varying $t$ over a $\poly(ksm)$ size set $T\subseteq \F$.

\paragraph{Proof of part(2):} The proof of the second part is similar. For some matrix tuple $(p_1,\ldots,p_n)\in \M_{km}(\F)$, let $d\leq 2sm -1$ such that $S^{\{d\}}$ is invertible at $\ubar{p}$ under the inclusion map $\iota : \M_m(\F)\rightarrow \M_{km}(\F)$ given by $\iota : a\rightarrow a\otimes I_k$. Let $M(t) = \r(tp_1 + q_1\otimes I_k, \ldots, tp_n + q_n\otimes I_k)$. As before, consider two interpretations of $M(t)$. Think of $M(t)$ as the evaluation of the generalized series $\r(\ubar{x}+\ubar{q})$ at $(tp_1, \ldots, tp_n)$ again under the inclusion map $\iota : \M_m(\F)\rightarrow \M_{km}(\F)$ given by $\iota : a\rightarrow a\otimes I_k$. We write $\det M(t) = t^{mkd} \det S^{\{d\}}(\ubar{p}) + M''(t)$ where $t$-degree of each term of the matrix $M''(t)$ is strictly more than $mkd$. Therefore, $\det M(t)$ is nonzero.

Interpret $M(t)$ as the evaluation of the rational formula $\r$ on $(tp_1 + q_1\otimes I_k, \ldots, tp_n + q_n\otimes I_k)$. Since $\r$ is a rational formula of size $s$, each entry of the matrix $M(t)$ is an element of the function field $\F(t)$. 
Again by Lemma \ref{lem:bivariate-degree}, the $t$-degrees of each numerator and denominator polynomial involved in the computation of $M(t)$ and $\det M(t)$ is also bounded by $\poly(ksm)$. 
Therefore, the final choice of the parameter $t$ should be such that it avoids the zeros of all such the numerator and denominator polynomials involved in the computation of $M(t)$ and $\det(M(t))$. This is clearly possible by varying $t$ over a $\poly(ksm)$ size set $T\subseteq \F$.

Final substitution is of the following form in both the cases:
\begin{equation}\label{eq-finalmatrix}
\{(\alpha p_1 + q_1\otimes I_k, \ldots, \alpha p_n + q_n\otimes I_k)\},
\end{equation}
for some suitably chosen $\alpha\in T\subseteq \F$.
\end{proof}


\subsection*{Strong hitting set for rational formulas of inversion height one}
We now show the construction of a strong hitting set for noncommutative rational formulas of inversion height one.

\begin{theorem}\label{thm:invertible-height-one}
Given $n,s$, we can construct a strong hitting set $\widetilde{\mathcal{H}}_1$ of size $(ns)^{O(\log ns)}$ over $\M^n_{d'}(K)$ for the class of noncommutative rational formulas $\r\in\Q\newbrak{x_1, \ldots, x_n}$ of size $s$ and of inversion height one. The parameter $d'$ is $\poly(n,s)$ and $K=\Q(\omega,z)$ is the extension field by adjoining a primitive root of unity $\omega$ of order $\ell$ where $\ell=\poly(n,s)$.       
\end{theorem}

\begin{proof}
Let $\r(\ubar{x})$ be a rational formula of inversion height one in $\Q\newbrak{\ubar{x}}$ of size $s$. Let $h_1, \ldots, h_k$ be all the sub-formulas input to the inverse gates in the rational formula for $\r$. Consider the noncommutative formula $h=h_1 h_2 \cdots h_k$ in $\Q\angle{\ubar{x}}$ which is of size at most $s$ and degree is also bounded by $s$. 

By Theorem \ref{thm:embedded-forbes-shpilka}, we construct a hitting set $\mathcal{H}_0$ in $D^n$ where $D = (K/F, \sigma, z)$ is a cyclic division algebra of index $\ell = \poly(n,s)$ for noncommutative ABPs in $\Q\angle{\ubar{x}}$ of width and layers at most $s$. 
Then there is a point $\ubar{q}\in \mathcal{H}_0$ such that $h(\ubar{q})$ is invertible and hence $\r(\ubar{q})$ is defined. 

Following Theorem~\ref{rit-connection}, if $\r(\ubar{x})$ is nonzero then $\r(\ubar{x} + \ubar{q})$ can be represented as a nonzero recognizable generalized series over $D$. Moreover, using the second part of Proposition~\ref{prop-gen-Sch}, to obtain a strong hitting set for $\r(\ubar{x})$, it suffices to find a strong hitting set of a generalized ABP over $D$ of width $r \leq 2s$ and degree $d \leq 2s\ell-1$. We now use the strong hitting set $\mathcal{H}_1$ in $\M^n_{d\ell}(K)$ (recall that $K = \Q(z,\omega)$ where $\omega$ is the primitive root of unity of order $\ell$)
for generalized ABPs of degree $d$ over $D$ (here $\ell$ is the index of $D$) obtained in Theorem~\ref{thm:gen-abp-hitting}. Inspecting the proof of Proposition~\ref{prop-gen-Sch}, we can now find a subset $T\subseteq \Q$ of size $\poly(n,s)$ and the final quasipolynomial-size hitting set is the following: 
\[
\widehat{\mathcal{H}}_1 = \{ \alpha\ubar{p} + \ubar{q}\otimes I_d  :  \ubar{p}\in\mathcal{H}_1, \ubar{q}\in\mathcal{H}_0, \alpha\in T\}\subseteq \M^n_{d\ell}(K).  
\]
\end{proof}

\subsection*{Hitting set for rational formulas of inversion height two}

We are now ready to prove our main theorem.\\

\maintheorem~~
 Let $\r(\ubar{x})$ be a rational formula of inversion height two in $\Q\newbrak{\ubar{x}}$ of size $s$. Let $\mathcal{F}$ be the collection of all those inverse gates in the formula such that for every $\mathfrak{g}\in \mathcal{F}$, the path from the root to $\mathfrak{g}$ does not contain any inverse gate. For each $\mathfrak{g}_i\in \mathcal{F}$, let $h_i$ be the sub-formula input to $\mathfrak{g}_i$.
Consider the formula $h=h_1 h_2 \cdots h_k$ which is of size at most $s$. 
 Clearly, $h$ is of inversion height one.
By Theorem \ref{thm:invertible-height-one}, we construct a strong hitting set $\widehat{\mathcal{H}_1}$ in $\M_d(K)$ where $d = \poly(n,s)$. Then there is a point $\ubar{q}\in \widehat{\mathcal{H}_1}$ such that $h(\ubar{q})$ is invertible and hence $\r(\ubar{q})$ is defined. 

Following Theorem~\ref{rit-connection}, if $\r(\ubar{x})$ is nonzero then $\r(\ubar{x} + \ubar{q})$ can be represented as a nonzero recognizable generalized series over $\M_d(K)$. Moreover, using the first part of the proof of Proposition~\ref{prop-gen-Sch}, to obtain a hitting set for $\r(\ubar{x})$, it suffices to find a hitting set for generalized ABP $B$ over $\M_d(K)$ of width $r \leq 2s$ and degree $\hat{d} \leq 2sd-1$, the degree-$\hat{d}$ truncated part of the generalized series $\r(\ubar{x}+\ubar{q})$. We recall the substitution map $\psi$ from Proposition \ref{prop-gen-Sch} and consider $\psi(B)$. Each entry of $\psi(B)$ is computable by a noncommutative ABP of width $2sd$ and degree $\hat{d}$ over $Z=\{z_{i,j,k'}\}$ variables.
Let $\mathcal{H}_{FS}\subseteq \M^{nd^2}_{\hat{d}+1}(K)$ be the hitting set for ABPs of width $2sd$ and of degree $\hat{d}$ over $nd^2$ many variables obtained from Theorem \ref{forbesshpilka}. We now define $\widetilde{\mathcal{H}}_{FS}\in \M^n_{d(\hat{d}+1)}(K)$ in the following way. For every matrix substitution in $\mathcal{H}_{FS}$, define a matrix substitution for each $x_{k'}$ as a $d(\hat{d}+1)$ matrix which can be thought of as a $d\times d$ block matrix whose $(i,j)^{th}$ block is the matrix substituted for $z_{i,j,k'}$ variable from $\H_{FS}$. It follows that $\H_{FS}$ is a hitting set of $B$ under the inclusion map $a\mapsto a\otimes I_{\hat{d}+1}$.

\begin{remark}
To see the reason that we use the inclusion map 
$a\mapsto a\otimes I_{\hat{d}+1}$, we give a simple illustrative example. Consider a generalized monomial $a_1 x_1 a_2 x_2 a_3$ where $a_1, a_2, a_3$ are $2\times 2$ matrices. Now the substitution map $\psi$ replaces the variables $x_1,x_2$ by $2\times 2$ symbolic matrices over noncommutative $Z$ variables. So the entries of the output $2\times 2$ matrix are noncommutative polynomials over $Z$ variables. Now substituting the $Z$ variables by $3\times 3$ matrices is equivalent to substituting $x_1, x_2$ by $6\times 6$ matrices putting the $3\times 3$ matrices in the corresponding blocks and evaluating it under the inclusion map that blows up the $2\times 2$ matrices $a_i :1\leq i\leq 3$ to $a_i\otimes I_3$. 
\end{remark}

Inspecting the proof of Proposition~\ref{prop-gen-Sch}, we can now find a subset $T\subseteq \Q$ of size $\poly(n,s)$ and the final quasipolynomial-size hitting set is the following: 
\[
\mathcal{H}_2 = \{ \alpha \ubar{p} + \ubar{q}\otimes I_{\hat{d} + 1} : \ubar{p}\in\widetilde{\mathcal{H}}_{FS}, \ubar{q}\in \widehat{\mathcal{H}}_1, \alpha\in T \}.
\]

Now we discuss how to obtain the hitting set over $\Q$ itself. 
We can think of $\omega$ and $z$ as place-holder variables $t_1, t_2$ of degree bounded by $\ell$. 
So, thinking $t_1,t_2$ just as indeterminates, for any nonzero rational formula $\r$, there exists a matrix tuple in the hitting set on which $\r$ evaluates to a nonzero matrix $M(t_1,t_2)$ of dimension $\poly(n,s)$ over $\Q(t_1,t_2)$. By Lemma \ref{lem:bivariate-degree}, each entry of $M(t_1,t_2)$ is a rational expression (in $t_1,t_2$) where the degrees of the numerator and  denominator polynomials are bounded by $\poly(s,n)$. Hence by the same argument sketched before,  we can vary the parameters $t_1, t_2$ over a sufficiently large set $\widetilde{T}\subseteq \Q$ of size $\poly(s,n)$ such that we avoid the roots of the numerator and denominator polynomials involved in the computation. This gives our final hitting set 
$\widetilde{\mathcal{H}}_2=\{ \ubar{q'}(\alpha_1, \alpha_2) : \ubar{q'}(\omega,z)\in
\mathcal{H}_2, (\alpha_1, \alpha_2)\in \widetilde{T}\times \widetilde{T}\}$. 
\qed

\section{Concluding Remarks}\label{sec:conclusion} 
In this paper we give a deterministic quasipolynomial-time algorithm to solve the identity testing of rational formulas of inversion height two in black-box model via a quasipolynomial-size hitting set construction. Can our technique be extended to obtain a quasipolynomial-size hitting set for higher inversion heights? From our proof technique it follows that if we have a quasipolynomial-size \emph{strong} hitting set for rational formulas of inversion height $h-1$, then we can bootstrap that to construct a hitting set for rational formulas of inversion height $h$. As shown in this paper, we are able to construct a strong hitting set only for rational formulas of inversion height one via the embedding of Forbes-Shpilka hitting set in a suitable division algebra. We conjecture it is possible to construct quasipolynomial-size hitting sets for rational formulas of any constant inversion height inside a division algebra of polynomially bounded index (with $2^{O(h)}$ as exponent for inversion height $h$), and we believe that generalized division algebras \cite{Jacobson96} could be useful for the construction.

\bibliographystyle{alpha}

\bibliography{ref2}

\end{document}